\documentclass[draft,prd,aps,showpacs]{revtex4}
\usepackage{amsmath,amssymb}

\newcommand{\be}{\begin{equation}}
\newcommand{\ee}{\end{equation}}
\newcommand{\ba}{\begin{eqnarray}}
\newcommand{\ea}{\end{eqnarray}}

\newcommand{\beq}{\begin{equation}}
\newcommand{\eeq}{\end{equation}}

\newcommand{\nn}{\nonumber}

\newcommand{\pu}{\bar\theta^{\dot \alpha}}

\newcommand{\bea}{\begin{eqnarray}}
\newcommand{\eea}{\end{eqnarray}}

\begin{document}

%
\title{BPS Analysis of  Gauge Field-Higgs Models in Non-Anticommutative Superspace}
\author{L.G.~Aldrovandi, D.H.~Correa,
F.A.~Schaposnik\footnote{F.A.S. is associated with CICBA.} and
G.A.~Silva\footnote{G.A.S. is associated with CONICET}}

\affiliation{Departamento\ de F\'{\i}sica, Universidad Nacional de
La Plata,
      C.C. 67, (1900) La Plata, Argentina.}

\begin{abstract}
 We extend the study of BPS equations in ${\cal
N}=1/2$ super Yang-Mills theory to the case of models with gauge
symmetry breaking. We first consider an Abelian gauge-Higgs
supersymmetric Lagrangian in $d=4$ dimensional Euclidean space
obtained by deforming ${\cal N} = 1$ superspace. The
supermultiplets include chiral and vector superfields and its
bosonic content coincides with that of the Abelian Higgs model
where vortex solutions to the BPS equation are known to exist in
the undeformed case. We also consider the $d=3$ dimensional
reduction of a non-Abelian $d=4$ deformed model and study its
deformed BPS equations, showing the existence of new monopole
solutions which depend on the deformation parameter.
\end{abstract}


\maketitle

\section{Introduction}

Non-anticommutative (NAC) theories recently attracted much
attention because of their relation with superstring effective
actions in backgrounds with constant graviphoton field strength
\cite{deBoer}-\cite{Ber1}. They can be constructed, within the
superfield formulation of SUSY theories, by introducing different
deformations in the odd superspace variables algebra
 \cite{Casal}-\cite{Araki}. As in ordinary noncommutative space, one
can introduce a Moyal star product to multiply superfields
entering in the construction of NAC Lagrangians. Depending on
whether one chooses  the supercovariant derivatives $D_\alpha$
 \cite{Ferrara1}
or the supersymmetric generators $Q_\alpha$ \cite{Seiberg} to
define such star product, one obtains a supersymmetric (but
chirality non-preserving)  theory or a partially supersymmetric
(but chirality preserving) one. Following this last approach,
Seiberg \cite{Seiberg} studied  ${\cal N}=1$ superspace and
constructed a super Yang-Mills Lagrangian in $d=4$ Euclidean space
which differs from the undeformed one in a polynomial in the
deformation parameter with terms containing fermion bilinear
products.  The resulting deformation reduces the supersymmetry of
the action from ${\cal N}=1$ to ${\cal N}=1/2$.

In order to study non-perturbative aspects of Seiberg's ${\cal
N}=1/2$ super Yang-Mills theory, instanton solutions were
constructed in \cite{Imaanpur}-\cite{billo}. As stressed in
\cite{Seiberg}, if one restricts the analysis to the purely
bosonic sector (putting fermions to zero) self-duality and
anti-self-duality equations are not modified. One can study
however how the bosonic equations get modified when fermions are
turned on. One possibility is to arrange the action functional
into
 perfect squares. Now, since the deformed action is in general complex,
the first order equations resulting from the vanishing of the
squares should be understood as corresponding to an enhancement of
the symmetry instead of leading to  a minimum of the action.
Interestingly enough, one finds that only the anti-self-duality
equations are modified in the deformed Super-Yang-Mills theory
\cite{Imaanpur}-\cite{britto}. Following a similar  approach, some
soliton  solutions in $d=2$ NAC theories were discussed
\cite{Abbaspur}-\cite{Chandra}.

As it is well-known,  vortex and monopole BPS equations can be
obtained by studying the supersymmetric extension of Abelian and
non-Abelian gauge theories coupled to Higgs scalars
\cite{DiV}-\cite{ENS}. It is the purpose of the present work to
extend this analysis to the case of NAC theories. We start in
section 2 by discussing deformed superspace and then consider a
gauge-Higgs supersymmetric Lagrangian in $d=4$ dimensional
Euclidean space obtained by deforming ${\cal N} = 1$ superspace.
The supermultiplets include chiral and vector superfields
(containing a complex scalar and a $U(1)$ gauge field
respectively) so that the bosonic content coincides with that of
the Abelian Higgs model. We show in section 3 that although new
terms arise due to the deformation, the resulting equations of
motion and supersymmetry transformations  show that no consistent
first order Bogomol'nyi equations arise except when fermions are
turned off (this eliminating the deformation effects). That is, in
contrast with what happens in the instanton case, one cannot find
deformed vortex configurations solutions. In order to make a
similar analysis for monopoles we discuss in Section 4 the $d=3$
dimensional reduction of a non-Abelian $d=4$ deformed model. In
this case one gets deformed BPS equations and new monopole
solutions which depend on the deformation parameter. We present a
discussion of our results in Section 5. We give in an Appendix
some conventions adopted in our calculations.

\section{Deformed Superspace}

We shall consider the deformation of 4 dimensional Euclidean
${\cal N} = 1$ superspace parametrized by superspace bosonic
coordinates $x^\mu$ and chiral and anti-chiral fermionic
coordinates $\theta^\alpha,\bar\theta^{\dot \alpha}$ as introduced
in \cite{Seiberg} \be
 \{\theta^\alpha, \theta ^\beta\} = C^{\alpha
 \beta} \; , \;\;\; \;\;\;
 \{\pu,\bar\theta^{\dot \beta}\} = 0 \; , \;\;\; \;\;\;
  \{\theta^\alpha,\bar\theta^{\dot \beta}\} = 0
 \label{uno}
\ee
Here $C^{\alpha \beta}$ are constant  elements of a symmetric
matrix. Defining chiral and anti-chiral coordinates according to
\ba
 y^\mu &=& x^\mu +i \theta \sigma^{\mu} \bar\theta\\
 \bar y^\mu &=& y^\mu -2i \theta \sigma^{\mu} \bar\theta
 \label{antichi}
\ea one imposes \cite{Seiberg}
\be
 [y^\mu,y^\nu] = [y^\mu, \theta^\alpha] = [y^\mu,\pu] = 0
 \label{dos}
\ee
and obtains as a consequence of (\ref{uno})-(\ref{dos}) \be
 [\bar y^\mu,\bar y^\nu] = 4\bar\theta\bar\theta \mathbf{C}^{\,\mu\nu}.
 \label{tres}
\ee where
$\mathbf{C}^{\,\mu\nu}=C^{\alpha\beta}(\sigma^{\mu\nu})_{\alpha\beta}$
is antisymmetric and antiselfdual (See the Appendix for
conventions on gamma matrices and spinors).

The non-anticommutative field theory in such a deformed superspace
can be defined in terms of superfields that are multiplied
according to the following  Moyal product \cite{Seiberg}
\be \Phi(y, \theta, \bar \theta) * \Psi (y, \theta, \bar \theta) =
\Phi (y, \theta, \bar \theta) \exp\left(-\frac{C^{\alpha\beta}}{2}
\frac{\overleftarrow{\partial}}{\partial \theta^\alpha} \frac{
\overrightarrow{\partial}}{\partial \theta^\beta} \right) \Psi(y,
\theta, \bar \theta) \label{moyprod} \ee
Supercharges and covariant derivatives in chiral coordinates take
the form \be Q_\alpha = \frac{\partial}{\partial \theta^\alpha} \;
, \;\;\;\;\; \;\;\;\;\; \;\;\;\;\; \;\;\;\;\; {\bar Q}_{\dot
\alpha} = - \frac{\partial}{\partial {\bar\theta}^{\dot\alpha}} +
2i \theta^\alpha \sigma^\mu_{\alpha\dot\alpha}
\frac{\partial}{\partial y^\mu}, \ee \be {D}_{ \alpha} =
\frac{\partial}{\partial {\theta}^{\alpha}} + 2i
\sigma^\mu_{\alpha\dot\alpha}{\bar \theta}^{\dot\alpha}
\frac{\partial}{\partial y^\mu}  \; , \;\;\; \;\; \;\;\;\;\;
\;\;\;\;\; \;\;\;\;\; {\bar D}_{\dot \alpha} = -
\frac{\partial}{\partial {\bar\theta}^{\dot \alpha}} \ee
The $D-D$ algebra is not modified by the deformation (\ref{uno})
as happens for the  $Q-D$ and $\bar Q-D$ algebra. Concerning the
supercharge algebra, it is modified according to
\ba
 \{{\bar Q}_{\dot \alpha},Q_\alpha \} &=&
 2i \sigma^\mu_{\alpha \dot \alpha} \frac{\partial}{\partial y^\mu} \\
 \{Q_\alpha,Q_\beta\} &=& 0\\
 \{{\bar Q}_{\dot \alpha}, {\bar Q}_{\dot \beta} \} &=& -4
 C^{\alpha\beta} \sigma^\mu_{\alpha \dot \alpha}\sigma^\nu_{\beta
 \dot \beta} \frac{\partial^2}{\partial y^\mu \partial y^\nu}
\ea
Then, only the subalgebra generated by $Q_\alpha$ is still
preserved and this defines the chiral ${\cal N} = 1/2$
supersymmetry algebra \cite{Seiberg}.

A chiral superfield $\Phi$ satisfying $\bar D_{\dot \alpha} \Phi =
0$ can be, as usual, written in the form
\be
 \Phi(y,\theta) = \phi(y) + \sqrt 2 \theta \psi(y) +
 \theta\theta F(y)
\ee As a consequence of (\ref{tres}) an ordering must be choosen
for the anti-chiral field $\bar \Phi (\bar y, \bar \theta)$, a
natural one is given by expressing it in terms of the chiral
variable $y^\mu$, it then takes the form
 \bea
 \bar \Phi(y-2i\theta\sigma\bar\theta,\bar\theta) &=& \bar\phi(y) +
\sqrt 2\bar\theta\bar\psi(y)
 -2i\theta\sigma^\mu\bar\theta\partial_\mu\bar\phi(y)
 +\bar\theta\bar\theta\left(\bar F(y)+
 i\sqrt{2}\theta\sigma^\mu\partial_\mu\bar\psi(y)
 +\theta\theta\partial^\mu\partial_\mu\bar\phi\right)
 \label{este}
\eea
Now, let us consider a vector superfield $V$ containing the gauge
field for a group $G$. We take $t^a$ as basis of the Lie algebra
satisfying $[t^a,t^b]=if^{abc} t^c$ and tr$(t^at^b)=\frac12
\delta^{ab}$. A gauge transformation acts as
\be
 \exp (-2gV) \to \exp (-2gV') = \exp (ig\bar \Lambda) * \exp
 (-2gV) * \exp (-ig \Lambda)
 \label{gt}
\ee where $\Lambda$ and $\bar\Lambda$ are chiral and anti-chiral
fields in the Lie algebra of $G$. In all the expressions above
exponentials are defined through their $*$-product expansion, \be
\exp (i\Omega) = 1 + i \Omega + \frac{i^2}{2} \Omega * \Omega +
\ldots \ee
For the chiral and anti-chiral superfield strengths, the standard
expressions hold, \bea
 W_\alpha &=& \frac{1}{8g} \bar D * \bar D * \exp(2gV) *
 D_\alpha * \exp(-2gV)\nonumber\\
 {\bar W}_{\dot \alpha} &=& -\frac{1}{8g} D * D * \exp(-2gV)* {\bar
D}_{\dot\alpha}* \exp(2gV) \eea transforming under gauge rotations
according to \be
 W_\alpha \to \exp (ig\Lambda) * W_\alpha * \exp
 (-ig\Lambda) \; , \;\;\;
 {\bar W}_{\dot\alpha}\to \exp (ig\bar
 \Lambda) *{\bar W}_{\dot \alpha}* \exp (-ig\bar\Lambda)
\ee Infinitesimally we have \be
 \delta W=ig[\Lambda,W]_*\,,~~~\delta \bar W=ig[\bar\Lambda,\bar W]_*.
\ee Since the commutator involves matricial and Moyal products, as
in standard noncommutative gauge theories one should consider
groups closing their Lie algebra generators under anticommutation.

We want to write  the vector superfield in the Wess-Zumino gauge.
As in ordinary superspace this is achieved by exploiting the gauge
freedom (\ref{gt}) to set  some of the components of  $V$  to
zero. In the generalization to non-anticommuta\-ti\-ve theory the
vector superfield $V$ in the Wess-Zumino gauge takes the form
\cite{Seiberg} \bea
 V(y,\theta.\bar\theta) &=& -\theta \sigma^\mu \bar \theta
 A_\mu(y)
 - i \bar\theta\bar\theta \theta^\alpha
 \left(\lambda_\alpha(y) -
 \frac{g}{2}\varepsilon_{\alpha\beta}C^{\beta\gamma}
 \sigma^\mu_{\gamma\dot\gamma}\{{\bar \lambda}^{\dot
 \gamma}(y),A_\mu(y)\} \right)
 + i\theta\theta\bar\theta \bar \lambda(y)
\nonumber
\\ &&
  + \frac{1}{2} \theta\theta\bar\theta\bar\theta
 (D(y) - i \partial_\mu A^\mu(y))
 \label{wz}
\eea
This leads to \ba
 V_*^2&\equiv&
 -\frac{1}{2}\,\bar\theta\bar\theta\left[ \theta\theta A_\mu
 A ^\mu + \mathbf{C}^{\mu\nu}A_{\mu}A_{\nu}
 -i\theta_\alpha C^{\alpha\beta}\sigma^\mu_{\beta\dot\beta}
[A_\mu,\bar\lambda^{\dot\beta}]
+\frac{1}{4}|\mathbf{C}|^2\bar\lambda\bar\lambda\right]\nn\\
 V_*^3&=&0
\ea
where $|\mathbf{C}|^2=\mathbf{C}^{\mu\nu}\mathbf{C}_{\mu\nu}$. One
can still perform gauge transformations preserving Wess-Zumino
gauge (\ref{wz}) through
\ba
\Lambda &=& \varphi(y)\\
\bar\Lambda &=&\varphi(y)
 -2i\theta\sigma^\mu\bar\theta\partial_\mu\varphi(y)
 +\theta\theta\bar\theta\bar\theta\partial^\mu\partial_\mu\varphi(y)
 - i g\bar\theta\bar\theta
\mathbf{C}^{\mu\nu}\{\partial_\mu\varphi(y),A_\nu(y)\} \nn\ea
In components this gauge transformation reads
\ba
 \delta A_\mu &=& {\cal D}_\mu \varphi\equiv\partial_\mu
\varphi-ig[A_\mu,\varphi]  \nn\\
 \delta \lambda_\alpha &=& -ig[\lambda_\alpha,\varphi] \nonumber\\
 \delta{\bar \lambda}_{\dot \alpha} &=& -ig[{\bar \lambda}_{\dot
\alpha},\varphi]\nonumber\\
\delta D &=& -ig[D,\varphi] \ea
 Chiral superfields charged under the gauge group transform
according to \be
 \Phi \to \exp (ig\Lambda) * \Phi \; ,
  \;\;\;\;\;
  \bar \Phi \to \bar \Phi * \exp (-ig \bar\Lambda)
 \ee
As in the case of the vector superfield in eq.(\ref{wz}), a
$C$-dependent term is needed in the parametrization of anti-chiral
matter superfields in order for the field components to have the
ordinary gauge transformation \cite{Araki}
 \ba
  \bar \Phi(\bar y, \bar\theta) &=& \bar\phi(\bar y)+\sqrt
2\bar\theta\bar\psi(\bar y)
  +\bar\theta\bar\theta\left(\vphantom{{a^2}^2}\bar F(\bar y)+
  2ig\mathbf{C}^{\mu\nu}\partial_\mu(\bar\phi(\bar y)A_\nu(\bar
y))+ {g}^2\mathbf{C}^{\mu\nu}\bar\phi(\bar y) A_\mu(\bar y)
A_\nu(\bar y)\right)
  \label{este2}
\ea
Then, written in components,  infinitesimal gauge transformations
read \bea
 \delta\phi = ig \varphi \phi \;\;\; && \;\;\; \delta\bar \phi =
 -ig  \bar \phi \varphi
 \nonumber\\
 \delta\psi =ig  \varphi \psi \;\;\; && \;\;\;  \delta\bar\psi =
 -ig \bar\psi  \varphi
 \nonumber\\
 \delta F = ig  \varphi F \;\;\;  && \;\;\; \delta \bar F =  -ig
 \bar F \varphi \label{transfor}
\eea
\section{Supersymmetric Maxwell-Higgs model in $d=4$ and
deformed vortices}

The $d=4$ deformed supersymmetric Maxwell-Higgs model is
constructed with the multiplets discussed in the previous section
as
\bea
 {\cal L}&=&\int\! d^2\theta d^2\bar\theta\,\,
 \left(\bar\Phi*\exp(-2gV)*\Phi + 2g v_0^2 V \right)
 +\frac{1}{4}\left(\int\! d^2\theta\, W*W\right.
 + \left. \int\! d^2\bar\theta\, \bar W*\bar W\right)
\eea
where all superfields are multiplied using the Moyal product
(\ref{moyprod}). A Fayet-Iliopoulos term has been included in
order to achieve spontaneous gauge symmetry breaking. In
components, the Lagrangian reads
\be
 {\cal L} = {\cal L}_b + {\cal L}_f \label{lagi}
\ee where
\bea
 {\cal L}_b &=& -\frac{1}{4}
 F_{\mu\nu} F^{\mu\nu} - \overline{D_\mu \phi} D^\mu \phi - g
 D(\bar \phi \phi   - v_0^2)  + \frac{1}{2} D^2 + \bar F F - {ig}
 \mathbf{C}^{\mu\nu}\bar \phi F_{\mu\nu} F
 \\
 {\cal L}_f &=& -i {\bar \lambda}\bar\sigma^\mu \partial_\mu \lambda -i
  {\bar \psi}\bar\sigma^\mu  {D}_\mu \psi
   -{i}{\sqrt{2}}g(\bar \phi \lambda \psi - \bar \psi \bar \lambda
\phi)
 +{i}g  \mathbf{C}^{\mu\nu}F_{\mu\nu}\bar \lambda \bar \lambda
 +  {\sqrt 2}g C^{\alpha \beta}\sigma^\mu_{\alpha \dot \alpha}
 \overline{D_\mu  \phi}\, \bar\lambda^{\dot \alpha}\psi_\beta
 -\frac{ g^2}{4}|\mathbf{C}|^2 \bar \phi\bar \lambda \bar \lambda F
 \label{lag}
\eea
Here  \bea D_\mu &=& \partial_\mu  - ig A_\mu
\nonumber\\
F_{\mu\nu} &=&\partial_\mu A_\nu - \partial_\nu A_\mu \eea
The transformation laws associated with the ${\cal N}=1/2$
surviving supersymmetry read
\bea
 && \delta \phi = \sqrt 2 \xi \psi \; , \;\;\;
 \delta \bar \phi =0 \nonumber\\
 && \delta \psi_\alpha = \sqrt 2 \xi_\alpha F \; , \;\;\;
 \delta \bar \psi_{\dot \alpha} =
 -i \sqrt 2 \overline{D_\mu\phi}
 \left(\xi \sigma^\mu\right)_{\dot \alpha}\nonumber\\
 && \delta F = 0 \nonumber\\
 &&   \delta \bar F = -i \sqrt 2\, \overline{D_\mu \psi} {\bar
 \sigma}^\mu \xi +2 ig\bar \phi \xi \lambda -2g
 \mathbf{C}^{\mu\nu}\left(
 \partial_\mu\left( \bar \phi \xi \sigma_\nu\bar \lambda\right)
 +{ig}\left( \bar \phi \xi \sigma_\nu\bar \lambda\right) A_\mu
 \right)\nonumber\\
 && \delta A_\mu = - i \bar \lambda {\bar \sigma}_\mu \xi  \;\;\;
 \rightarrow
 \;\;\;
 \delta F_{\mu\nu} = -i (\partial_\mu \bar \lambda \bar \sigma_\nu
 -
 \partial_\nu \bar \lambda \bar \sigma_\mu)\xi\nonumber\\
 && \delta \lambda_\alpha  = i\xi_\alpha D +
 \left(\sigma^{\mu\nu}\xi\right)_\alpha \left( F_{\mu\nu}
 -ig\, \mathbf{C}_{\mu\nu}\bar \lambda \bar \lambda
 \right)\nonumber \\
 && \delta \bar \lambda_\alpha = 0 \; , \;\;\;
 \delta D =-\xi\sigma^\mu \partial_\mu \bar \lambda
 \label{bogo}
\eea
The second order equations of motion associated to the Lagrangian
(\ref{lag}) are
\bea
 \partial_\mu F^{\mu\nu} &=& ig\left(\bar \phi D^\nu \phi -
  \phi\overline{ D^\nu \phi}\right) +g \bar \psi {\bar \sigma}^\nu
\psi+2i g \mathbf{C}^{\mu\nu}
  \partial_\mu (\bar\lambda\bar\lambda - \bar \phi F) -i{g^2} {\sqrt 2}
  C^{\alpha \beta}\sigma^\nu_{\alpha \dot \alpha}  \bar \phi
 \bar \lambda^{\dot \alpha}\psi_\beta  \label{max}\\
 D^\mu D_\mu \phi &=& g \phi D + {i}g{\sqrt 2}\lambda \psi+
ig\mathbf{C}^{\mu\nu}
 F_{\mu\nu} F  + {\sqrt 2}g
 C^{\alpha\beta} \sigma^\mu_{  \alpha \dot \alpha}D_\mu(\bar
 \lambda^{\dot \alpha} \psi_\beta)
  +g^2 \frac{|\mathbf{C}|^2}{4} \bar \lambda
 \bar \lambda F\label{higgs}\\
 \overline {D^\mu}\; \overline {D_\mu \phi} &=& g \bar \phi D -
{i}g{\sqrt 2}
   \bar \lambda\bar\psi\label{higgsb}\\
  F = 0 & , &
  \bar F = {i}g\bar \phi\, \mathbf{C}^{\mu\nu} F_{\mu\nu}
 +g^2\frac{|\mathbf{C}|^2}{4} \bar \phi\,\bar \lambda\bar \lambda
\label{FFFF}\\
 D &=& g (\bar \phi \phi - v_0^2) \label{DDD}\\
 (\sigma^\mu\partial_\mu\bar \lambda)_\alpha &=& -{\sqrt 2}g\bar \phi
 \psi_\alpha
 \label{bbb}\\
 (\bar\sigma^\mu \partial_\mu   \lambda)_{\dot \alpha} &=&
 {\sqrt{2}}g\phi\bar \psi_{\dot\alpha}   +2g
\mathbf{C}^{\mu\nu}F_{\mu\nu}\bar
 \lambda_{\dot\alpha} + i{\sqrt 2}g \overline{D_\mu \phi}\,\psi_\alpha
C^{\alpha\beta}
 \sigma^\mu_{\beta\dot\alpha}+ig^2 \frac{|\mathbf{C}|^2}{2}F \bar \phi  \bar
\lambda_{\dot\alpha}
 \label{lll}\\
 (\bar\sigma^\mu  {D}_\mu \psi)_{\dot\alpha} &=&{\sqrt{2}}g \phi  \bar
\lambda_{\dot\alpha}  \label{ppp}\\
 \left(\sigma^\mu \overline {D_\mu  \psi}\right)_\beta &=&
 -{\sqrt{2}}g\bar \phi \lambda_\beta
 -{i}{\sqrt 2}g \overline{D_\mu  \phi}  \bar \lambda_{\dot \alpha}
 {\bar \sigma}^{\mu\,\dot \alpha \alpha}
 C_{\alpha \beta}
 \label{ecuaciones}
\eea
In connection with gauge symmetry breaking it is interesting to
look for constant solutions to eqs.(\ref{max})-(\ref{ecuaciones})
to see whether the usual Higgs vacuum ($\phi =v_0$, $A_\mu
=\partial_\mu \Lambda$) is modified by the deformation. In
particular, one could think that the presence of new $C$-dependent
terms could lead to  gauge symmetry breaking even when $v_0^2 =
0$. This possibility is suggested by the supersymmetry  variation
of $\lambda$ which exhibits a term proportional to $C \bar\lambda
\bar \lambda$, which could play the role that $v_0^2$ does in the
normal case. Now, for constant fields,  the only equation
involving the deformation parameter $C$ is (\ref{max})
\be
 \bar \psi {\bar \sigma}^\nu \psi =  ig{\sqrt 2}
 C^{\alpha \beta}\sigma^\nu_{\alpha \dot \alpha}  \bar \phi
 \bar \lambda^{\dot \alpha}\psi_\beta \label{36}
\ee
In order to have a non-trivial $C$ contribution we need
$\bar\phi,\psi$ and $\bar\lambda$ to be non-vanishing constants.
However this is not possible in view of eq.(\ref{bbb}). We then
conclude that there is no non-trivial symmetry breaking mechanism
apart from that originated by the standard Fayet-Iliopoulos term.

In the $d=4$  super Yang-Mills theory case, instanton
configurations were constructed by solving a deformed version of
the first order self-duality equations
\cite{Imaanpur}-\cite{britto}. The deformation was originated by
the presence of fermionic zero modes. One could expect that in the
present case, deformed Nielsen-Olesen vortex configurations could
be obtained by solving some deformed first order Bogomol'nyi (BPS)
equations. To this end, let us restrict fields, from here on, to
the $x^1,x^2$ plane and make $A_3=A_4 =0$. Moreover, we shall
consider for simplicity   that the only nonvanishing
$\mathbf{C}^{\mu\nu}$ components are $\mathbf{C}_{12} = -
\mathbf{C}_{34}$.

In the undeformed case, BPS equations can be obtained from the
vanishing of the supersymmetry transformations for fermions, once
the auxiliary fields are put on-shell. Nontrivial solutions to
these equations are invariant under 1/2 of the original
supersymmetries. Let us then analyze the ${\cal N}=1/2$ surviving
supersymmetry variations (\ref{bogo}). There are two possibilities
for making the supersymmetry variations of the fermionic and the
auxiliary fields vanish: either $\xi_1= 0$ and the following first
order equations hold (``anti-self-dual case'')
\bea
 && F_{12} = g(\bar \phi \phi - v_0^2)
 -  {i}  \mathbf{C}_{12}\bar\lambda\bar\lambda\label{bpsa}\\
 && \overline{D_1 \phi} + i \overline{D_2 \phi} = 0
 \label{bpsb}\\
 && \sqrt 2 \left( {\bar D_1} - i { \bar D_2}\right)\bar \psi_{\dot
 1} + \bar \phi \lambda^2 -\mathbf{C}_{12}\bar \lambda_{\dot 1} \left(
 \overline{D_1\phi} - i\overline{D_2\phi}\right) = 0 \label{bpsc}\\
 &&
 (\partial_1 - i \partial_2)\bar \lambda^{\dot 2}  = 0 \; , \;\;\; F =
0  \label{bps} \eea
or $\xi_2= 0$ and the first order equations take the form
(``self-dual case''),
\bea
 && F_{12} = -g (\bar \phi \phi - v_0^2)
 -  {i}  \mathbf{C}_{12}\bar\lambda\bar\lambda\label{bpsbb}\\
 && \overline{D_1 \phi} - i \overline{D_2 \phi} = 0\label{bpscc}\\
 && \sqrt 2 \left(\bar D_1 + i \bar D_2\right)\bar \psi_{\dot 2} +
 \bar \phi \lambda^1 +\mathbf{C}_{12}\bar \lambda_{\dot 2} \left(
 \overline{D_1\phi} + i\overline{D_2\phi}\right) = 0 \label{bps2}\\
 && (\partial_1 + i \partial_2)\bar \lambda^{\dot 1}  = 0\label{bpsaa}
\; , \;\;\;
 F = 0
\eea
At this point, an important difference  with respect to the
Yang-Mills deformed case should be stressed. In the latter, for
anti-selfdual configurations, fermions  are invariant under the
whole ${\cal N}=1/2$ surviving symmetry while for selfdual
configurations they are not. In the present case, according to
(\ref{bpsa})-(\ref{bpsaa}) both for selfdual and anti-selfdual
configurations fermions would be  invariant under $1/2$ of the
${\cal N}=1/2$ supersymmetry which survived the deformation.

Let us discuss, for definiteness, the self-dual case
(eqs.(\ref{bpsa})-(\ref{bps}), the anti-self-dual one goes the
same). Compatibility of eq.(\ref{bps}) for $\bar\lambda^{\dot 2}$
with equation of motion (\ref{bbb}) implies that $\psi_1 = 0$. But
this in turn implies, because of eq.(\ref{ppp}), that $\bar
\lambda^{\dot 2} = 0$, so that finally $\bar \lambda \bar \lambda
= 0$: any effect from deformation is finally washed out.

In brief,  on the one hand one necessarily has to keep $\bar
\lambda \bar \lambda \ne 0$ in order to discover new features in
the deformed model. On the other hand, the deformed first order
BPS equations obtained from the vanishing of supersymmetry
transformations are not compatible with the equations of motion,
except if some fermionic fields vanish turning
 the deformed BPS equations  into the undeformed
(ordinary) ones.

The previous results can be also understood by noting that in fact
the deformed Lagrangian cannot be arranged as a sum of perfect
squares whose vanishing lead to  deformed first order equations
(\ref{bpsa})-(\ref{bps}), as one can do in the undeformed case.
Indeed,  one cannot reproduce Lagrangian (\ref{lag}) from, among
others, a square term of the form
\be
  \left(F_{12} - D - i \mathbf{C}_{12} \bar \lambda \bar\lambda
\right)^2 \ee
since a term of the form $iD  \mathbf{C}_{12} \bar \lambda
\bar\lambda$ is lacking in eq.(\ref{lag}). This again should be
contrasted with the case of deformed Yang-Mills theory, where
Lagrangian can be written as  squares of $C$-deformed self-duality
equations.

One can consider  the possibility of finding $C$-dependent
solutions by directly  analyzing  the equations of motion
(restricted to the $x^1,x^2$ plane and with $A_3=A_4=0$). However,
the set of coupled nonlinear equations lead to  very complicated
constraints. For example, Maxwell equations (\ref{max}) for
$\nu=3,4$ require
\be
 \bar \psi {\bar \sigma}^\nu \psi =  ig{\sqrt 2}
 C^{\alpha \beta}\sigma^\nu_{\alpha \dot \alpha}  \bar \phi
 \bar \lambda^{\dot \alpha}\psi_\beta \, , \;\;\;\;\; {\rm for}\;
\nu=3,4 \ee
and for non-vanishing $\psi_\alpha$ this leads to the  constraints
\bea
 \bar\psi^{\dot 1}&=& +\sqrt 2 i g C^{12} \bar\phi\bar\lambda^{\dot
1}\nn\\
 \bar\psi^{\dot 2}&=& -\sqrt 2 i g C^{12} \bar\phi\bar\lambda^{\dot 2}
 \label{con}
\eea
This in turn implies $\bar\psi\bar\lambda=0$. We were not able to
establish the compatibility of this result with the equation of
motion for $\bar \psi$ and although non-trivial $C$-dependent
vortex solution cannot be {\it a priori} excluded, it seems
extremely difficult to fulfill all the resulting constraints.

\section{Supersymmetric $U(2)$ Yang-Mills-Higgs model
in $d=3$ and deformed monopoles}

In this section we shall consider a deformed $d=3$ supersymmetric
$U(2)$ gauge theory coupled to scalars in order to analyse
possible modifications, induced by the deformation, on the BPS
(first order) monopole equations. To this end, we start from a
deformed supersymmetric $d=4$ Yang-Mills theory and proceed to a
dimensional reduction in which the $A_4$ component of the gauge
field is identified with a Higgs field. The $d=4$ Lagrangian in
term of superfields reads
\bea
 {\cal L}&=& \frac{1}{2}{\rm tr}
 \left(\int\! d^2\theta\, W*W  + \int\! d^2\bar\theta\,
 \bar W*\bar W\right)
\eea
To write the Lagragian in components, we use Dirac spinors and a
generic $\Gamma$ matrices representation,
\bea
 {\cal L}  &=& {\rm tr}\left( -\frac{1}{2}  F_{\mu\nu} F^{\mu\nu}
 -i  {\Lambda}^{\cal C} \Gamma^\mu {\cal D}_\mu \Lambda
 +2ig \mathbf{C}^{\mu\nu}F_{\mu\nu}\Lambda^{\cal C} P_- \Lambda
 + g^2|\mathbf{C}|^2   \left(\Lambda^{\cal C}
P_-\Lambda
 \right)^2 + D^2 \vphantom{\frac 12}\right) \label{lagna}
\eea
where
\bea
 F_{\mu \nu} &=& \partial_\mu A_\nu - \partial_\nu A_\mu -
 ig[A_\mu,A_\nu]\nonumber\\
 {\cal D}_\mu \Lambda &=& \partial_\mu \Lambda -
 ig[A_\mu,\Lambda]
\eea
Here we write $A_\mu = A_\mu^a t^a$ with $t^a$ the hermitic
generators normalized according to ${\rm tr\,} t^at^b =
 \frac{1}{2} \delta^{ab}$. For the present $U(2)$ case
 $t^a = \sigma^a/2$ ($a=1,2,3$) and $t^4= I/2$.

The equations of motion derived from Lagrangian (\ref{lagna}) are
\bea
 {\cal D}_\mu F^{\mu\nu}& =  &2i g\,
 \mathbf{C}^{\mu\nu}{\cal D}_\mu \left(\Lambda^{\cal C} P_-
\Lambda\right)
 + \frac{ g}{2}\{\Lambda^{\cal C}_\rho, \left(\Gamma^\nu
{\Lambda}\right)^\rho\}
 \nonumber\\
 \Gamma^\mu D _\mu   \Lambda
 &=&  g \,\mathbf{C}_{\mu\nu}\left\{F^{\mu\nu}
 -ig \mathbf{C}^{\mu\nu}  \Lambda^{\cal C}
P_-\Lambda,P_-\Lambda\right\}
 \nonumber\\
 D \vphantom{\Sigma^\mu} &=& 0
\eea
In the dimensional reduction, a vector field  in $d=4$ becomes a
vector field and a scalar field in $d=3$
\be
 A_\mu \to A_i \, , \;  \phi
\ee
where $\phi = \phi^a t^a $ ($a=1,2,3,4$) will play the role of a
Higgs field in the adjoint in $d=3$. Concerning fermions, the
$d=4$ Dirac  spinor $\Lambda$ reduces to two $d=3$ Dirac spinors
$\Lambda_1$ and $\Lambda_2$,
\be
 \Lambda = \left(
 \begin{array}{c}
 \Lambda^1\\
 \Lambda^2
 \end{array}\right) \to \Lambda^1 \, , \Lambda^2
\ee
For later convenience, we redefine $d=3$ fermions in the form \be
 \eta = \frac{1}{\sqrt 2}(\Lambda^1 + i \Lambda^2) \, , \;\;\;
 \chi= \frac{1}{\sqrt 2}(\Lambda^1 - i \Lambda^2)
\ee The dimensionally reduced $d=3$ Lagrangian then reads \bea
 {\cal L}  &=& {\rm tr}\left( - \frac{1}{2} F_{ij} F^{ij}
 - D^i\phi D_i\phi - 2i \chi^{\cal C}\gamma^i D_i\eta
 + 2 g \chi^{\cal C} [\phi,\eta]+D^2\right.
 \nonumber\\
 && \left. \;\;\;\;\; + 2 i g \mathbf{C}^{ij}\left(
 F_{ij} + \varepsilon_{ijk} D^k\phi \right)\chi^{\cal C}
 \chi+ 2{ g^2}\mathbf{C}^{ij}\mathbf{C}_{ij} \left(\chi^{\cal C}
\chi\right)^2
 \right)
 \label{lagui}
\eea
Here the Majorana conjugates should be computed by using ${\cal
C}_3$. The equations of motion for the bosonic fields read \bea
  && D_i \left(F^{ij} + ig \mathbf{C}^{ij} \chi^\mathcal{C}\chi\right)
=
 i{g} [D^j\phi
+ig\varepsilon^{jkl}\mathbf{C}_{kl}\chi^\mathcal{C}\chi,\phi]
 -
  \frac{g}{2}\eta^\mathcal{C}\gamma^i \chi
  \label{ecu0}\\
  && D^iD_i \phi = i g \varepsilon^{ijk} \mathbf{C}_{jk}
D_i\left(\chi^\mathcal{C}
  \chi\right)
  +\frac{g}{2}
  [\chi^\mathcal{C},\eta]  \label{ecu}
\eea
Concerning fermions, \bea
 i \gamma^i D_i \chi - g [\phi,\chi] &=&   0
 \label{lam2}
 \\
 i\gamma^iD_i \eta + g[\phi,\eta]
 &=&
  ig
 \mathbf{C}^{ij}\left\{\chi, F_{ij} - \varepsilon_{ijk} D^k\phi
  - 2ig \mathbf{C}_{ij} \chi^\mathcal{C}\chi)\right\} \label{motion}
\eea
The $d=3$ infinitesimal transformations associated with the
supersymmetry read \bea
 \delta \eta &=& - \gamma^i \xi
\left(
 \frac{1}{2}\varepsilon_{ijk}(F^{jk} -2
ig\mathbf{C}^{jk}\chi^\mathcal{C} \chi)
 +  D_i\phi \right)
 + i D \xi \label{laque}\\
 \delta\chi &=& 0 \\
 \delta D &=& - \xi^{\mathcal{C}} (\gamma^i D_i \chi +
 i g [\phi,\chi])
 \label{laque2}
 \\
 \delta F_{ij} &=&
 -{i} \xi^{\mathcal{C}}\left(\gamma_iD_j\chi  -
 \gamma_jD_i\chi
 \right)
 \\
\delta D_i \phi &=&
 -\xi^{\mathcal{C}}\left(iD_i \chi^\mathcal{C}+
 [\phi,\chi^\mathcal{C}] \gamma^i \right)
 \eea
Let us write variations (\ref{laque}),(\ref{laque2}) in the form
\bea
 \delta \eta^\alpha &=& {i} \xi_\beta L^{\alpha\beta} \nonumber\\
\delta D &=& {}\xi^{\mathcal{C}}_\alpha L^\alpha \eea with
$L_{\alpha\beta}$ and $L_\alpha$ appropriately defined.  We have
factored out $g$ so that a rescaling of all fields  $A_i, \phi, D,
\eta,\chi  \to g A _i, g\phi, g D, g\eta,g\chi$,
  renders the new
$L_{\alpha\beta}$ and $L_\alpha$ $g$-independent. Lagrangian
(\ref{lagui}) can be rewritten in the form \be
 {\cal L} = \frac{1}{ g^2}{\rm tr}
 \left( - L^{\alpha}_{\;\beta}L^{\beta}_{\;\alpha} +
 \eta^{\mathcal{C}}_\alpha L^\alpha\right) -\frac{1}{g^2}
 \varepsilon^{ijk}{\rm tr}
D_i\phi F_{jk} \ee
Then, in the $g^2 \to 0$ limit, dynamics is governed by
configurations which make  the supersymmetry variations associated
to $L_{\alpha\beta}$ and $L_\alpha$  vanish. That is,
configurations satisfying  the following first order equations,
\bea
 2 D_i\phi &=&
 \varepsilon_{ijk}(F^{jk} -2 i  \mathbf{C}^{jk}\chi^\mathcal{C}\chi)
 \label{nos}\\
 D&=&0 \label{di}
\eea
Concerning the vanishing for the auxiliary field variation,
 \be
 \gamma^i D_i \chi  +  i [\phi,\chi] =0
 \label{tresi}
\ee it just coincides with the equation of motion for $\chi$,
eq.(\ref{lam2}).

Arranging Lagrangian (\ref{lagui}) into perfect squares one can
see that whenever first order equations(\ref{di})-(\ref{tresi})
are satisfied, the action coincides with the topological
(magnetic) charge. Indeed, starting from (\ref{lagui}) one can
rewrite the corresponding action in the form
\bea S  &=& - \frac{1}{g^2} {\rm tr}\int d^3x \left(
 \left (\frac{1}{2}\varepsilon_{ijk}(F^{jk} -2
  i  \mathbf{C}^{jk}\chi^\mathcal{C} \chi) -
 D_i\phi \right )^2  -  D^2
  +{2 i} \eta^\mathcal{C} \left(\gamma^i D_i
 \chi  +   i [\phi,\chi] \right)
 \vphantom{\frac 12}\right) - \frac{1}{g^2}Q_M
 \label{laguic}
\eea
where $Q_M$ is a surface term related to the topological charge
\be
 Q_M = {\rm tr}\int dS_i \varepsilon^{ijk}
 F_{jk}\phi
\ee
Note that although we have managed to arrange the action in the
form (\ref{laguic}), we cannot ensure that configurations
satisfying eqs.(\ref{di})-(\ref{tresi}) lead to a bound for the
action given by the topological charge. This is because the
perfect square in the action is not positive definite since
$\mathbf{C}_{ij}$ is in general complex and $\chi$ in Euclidean
space is a Dirac spinor. One can easily see however that any field
configuration satisfying (\ref{di}), (\ref{tresi}) and $\eta = 0$
verifies the equations of motion (\ref{ecu0})-(\ref{motion}).
Eq.(\ref{nos}) can then be seen as the deformed extension of the
antiselfdual (BPS) equation for the Yang-Mills-Higgs system, the
analogous to the deformed Bogomol'nyi
eqs.(\ref{bpsb})-(\ref{bpsc}) for the Abelian-Higgs model.

~

Let us study solutions to equations (\ref{di})-(\ref{tresi}).
Evidently, the configuration \bea
 && \phi^a(x) = \phi^{PS\,a}(x) =
 \frac{x^a}{r^2} (\mu r \coth(\mu r) - 1) =
 \frac{x^a}{r} f(r) \,, \;\;\; a=1,2,3
 \nonumber\\
 && A_i^a(x) = A_i^{PS\,a}(x) = \varepsilon^{aij} \frac{x_j}{r^2}
 \left(1 - \frac{\mu r}{\sinh(\mu r)}\right) =
 \varepsilon^{aij} \frac{x_j}{r} (1-K)\,, \; a=1,2,3 \nonumber\\
 && \phi^{4}(x) = 0 \; , \;\;\; A_i^{4}(x) = 0  \; , \;\;\;
 \chi = 0
 \label{well}
\eea
where $\phi^{PS}$ and $A_i^{PS}$ are the well-honnored
Prasad-Sommerfield \cite{PS} monopole $SU(2)$ solutions with $\mu$
a constant with mass dimensions solves the first order system. The
effects of deformation should arise only if the fermion field
$\chi \ne 0$. As done in \cite{Imaanpur}-\cite{britto} for the
instanton case, we shall look for such solutions recursively,
starting from (\ref{well}) and writing \bea
 A_i^a(x) &=&  A_i^{PS\,a}(x) + A_i^{(1)\,a}(x) + \ldots \nonumber\\
 A_i^4(x) &=&    A_i^{(0)\,4}(x) + \ldots \nonumber\\
 \phi^a(x) &=&   \phi^{PS\,a}(x) + \phi^{(1)\,a}(x) \ldots \nonumber\\
 \phi^4(x) &=&     \phi^{(0)\,4}(x) +\ldots \nonumber\\
 \chi &=& \chi^{(0)} + \ldots
\eea Function $\chi^{(0)}$ can be obtained by solving
eq.(\ref{tresi}) in the background of a Prasad-Sommerfield
monopole. The answer is \be
 \chi^{(0)}  = D_i\phi^{PS} \gamma^i \zeta
\ee with $\zeta$ a constant spinor. One has now to insert this
solution in eq.(\ref{nos}) in order to compute the first order
corrections to the gauge and scalar fields. As in the instanton
case  the bilinear $\chi\chi$ is antisymmetric in the $U(2)$
indices and then the $\mathbf{C}_{\mu\nu}$ perturbation in
(\ref{bogo}) only affects the $U(1)$ subgroup. Then,  $SU(2)$
components of   the gauge and scalar fields  corrections vanish, $
A_i^{(1)\,a}(x)
 = 0$,
$\phi^{(1)\,a}(x) = 0$, $a=1,2,3$. Concerning the $U(1)$ sector,
one has to solve, for the first order correction, the equation

\be
 \varepsilon_{ijk} \partial_i\phi^{(0)4} + F^{(0)4}_{jk} =
- i \mathbf{ C}_{jk}\left(\chi^{(0)C}
\chi^{(0)}\right)^{\!\!\{ab\}} \varepsilon_{ab} \equiv
\mathbf{C}_{jk}J(x) \label{di0} \ee
with $\{ab\}$ indicating anti-symmetrization in $SU(2)$ indices.
Taking the derivative in both sides one gets for the gauge field
(taken in the Lorentz gauge) \be
 \nabla^2 \!A_k^{(0)4} =\mathbf{C}_{jk} \partial_j J
\label{nab} \ee Writing the gauge field in terms of a potential
$\Phi$ \be A_k^{(0)4}(x) = \mathbf{C}_{jk}\partial_j \Phi(x)
\label{aaa} \ee the  problem reduces to \be
\mathbf{C}_{jk}\partial_j\left(\nabla^2 \! \Phi - J\right) = 0
\label{edos} \ee with \be
 J = - i \left(\chi^{(0)\mathcal{C}}
 \chi^{(0)}\right)^{\!\!\{ab\}} \varepsilon_{ab}
 = - i D^i\phi^{a PS} D^i\phi^{a PS} \zeta^{\cal C} \zeta
\ee After some calculation one finds that the source $J$ takes the
form \be
 J = -i\left( \left(\frac{df}{dr}\right)^2 + \frac{1}{r^2} f^2
K^2\right)
 \zeta^{\cal C} \zeta
\ee With $f$ and $K$ as given in (\ref{well}) one finally has \be
 J = -  \left( \frac{1}{(\mu r)^4} + \frac{2}{\sinh^2(\mu r)}\left(
 1  -2 \frac{\coth(\mu r)}{\mu r}   + \frac{3}{2\sinh^2(\mu r)}
 \right)
 \right){i \zeta^{\cal C} \zeta}
\ee
A solution of eq.(\ref{edos}) can be obtained by solving the
Poisson equation \be
 \nabla^2 \! \Phi = J \label{poi}
\ee

Then, inserting the solution for $A_k^{(0)4}$ in eq.(\ref{di0})
one finds the solution for $\phi^{(0)4}$.

Since the only correction to the Prasad-Sommerfield solution was
the new $U(1)$ components $A_i^{(0)4}$ and $\phi^{(0) 4}$, the
zero mode equation for $\chi$ is not modified and hence the next
correction $\chi^{(1)} = 0$. Finally all higher order
 corrections both for bosonic and fermionic fields vanish.

\section{Summary and discussion}

The connection between self-dual or BPS equations and  ${\cal
N}=1$ and ${\cal N}=2$  supersymmetry  is by now well understood.
In this context, studying ${\cal N}=1/2$ supersymmetric models
allows to gain some control on relevant aspects of a kind of
interpolation towards the ${\cal N}=0$ model. An analysis of
instantons solutions in ${\cal N}=1/2$ super-Yang-Mills theory was
started in \cite{Seiberg} and advances on this issue were reported
in \cite{Imaanpur}-\cite{britto}. In this paper we have extended
the analysis to the case of solitons and instantons in deformed
supersymmetric theories with gauge  symmetry breaking. As in the
pure super-Yang-Mills case, the effect of deformation manifests at
the level of the gauge field-Higgs Lagrangians  through the
occurrence of a finite number of polynomial terms containing
fermion bilinears, both for the Abelian and non-Abelian models.
This modifies the surviving supersymmetry transformation law for
the gaugino and, consequently, the first order (``BPS'') equations
obtained when one imposes such transformations to vanish.

In the undeformed case,  the solution to the first order BPS
equations correspond to a bound for the action  as can be easily
seen by writing the (real) action or energy as a sum of perfect
squares plus a topological term-the bound. One can still try to
write the deformed action in that way but, being the action in
general complex, it has no sense to do this looking for for a
bound. This has been done for $\mathcal{N}=1/2$ super-Yang-Mills
where it was confirmed   that configurations satisfying the first
order equations arising from the vanishing of SUSY transformations
reduce the action to a topological charge \cite{Seiberg},
\cite{Imaanpur}-\cite{britto}. Concerning the deformed $d=3$
Yang-Mills-Higgs theory, we have shown here that the same can be
done. In contrast, this cannot be achieved for the
$\mathcal{N}=1/2$ supersymmetric Maxwell-Higgs action.

 Solutions to the first order BPS equations  can in principle be associated
with self-dual and anti-self-dual configurations which will be in
general differently affected by the deformation. It has been shown
in the instanton case \cite{Imaanpur}-\cite{britto}
 that antiselfdual configurations
are invariant under the whole ${\cal N}=1/2$ surviving symmetry
while selfdual configurations are not. We have shown that the same
happens in the monopole case.

In summary, we have shown that no deformed vortex solutions can be
found from the first order system except those where all fermions
are equal to zero, which reduce to the ordinary Nielsen-Olesen
vortices. In the $\mathcal{N}=1/2$ supersymmetric $d=3$
Yang-Mills-Higgs case, for which the deformed (complex) action can
be written as a sum of squares plus a topological charge,
solutions to the first order equations arising from the vanishing
of the gaugino supersymmetry variation can be found and they
correspond to anti-monopole configurations deformed by the
non-anticommutativity. We have analyzed these solutions using an
iterative process with the deformation parameter $C^{\alpha\beta}$
playing the  perturbation parameter. Due to the Grassmann nature
of the perturbing fermion field, this iterative procedure stops
and in this sense an exact deformed monopole solution can be
constructed.

Let us end by mentioning  that a connection between the kind of
deformation we have discussed and the spectral degeneracy of
conventional $\mathcal{N}=1$ SUSY gluodynamics has been recently
discussed in \cite{GS}. Remarkably, the analysis in this work
suggests that ${\cal N} = 1/2$  supersymmetry remains valid for
 coordinate dependent $C^{\alpha\beta}$. An analysis of such kind of
deformations
 in supersymmetric gauge field-Higgs models, extending the one
presented here would then be of interest. We hope to
 report on this issue in future work.

\vspace{1.5 cm}

\noindent\underline{Acknowledgements}: We wish to thank R. Britto
for helpful comments. This work  was partially supported by UNLP,
CICBA, CONICET, ANPCYT (PICT grant 03-05179)  D.H.C was partially
supported by Fundaci\'on Antorchas and CONICET.

\appendix
\section{Appendix }

 \noindent\underline{The chiral representation}

\vspace{0.3 cm}

In sections 2 and 3 we closely follow Wess-Bagger conventions. An
important point about $d=4$ Euclidean space is that no h.c.
relation exist between $\theta^\alpha$ and $\pu$.
 \ba
 \Gamma^\mu&=&\left( \begin{array}{cc}
   0 & \sigma^\mu \\
   \bar\sigma^\mu & 0 \\
 \end{array}\right),~~~
 \sigma^\mu=\left(i\vec\sigma,-1\right),~~
 \bar\sigma^\mu=\left(i\vec\sigma,1\right),~~
 {\rm tr}\,\sigma^\mu\bar\sigma^\nu=-2\delta^{\mu\nu}\\
 \Gamma_5&=&\Gamma^1\cdots\Gamma^4=\left( \begin{array}{cc}
   1 & 0 \\
   0 & -1 \\
 \end{array}\right),~~(\Gamma_5)^2=1\\
 {\cal C}&=&\Gamma^3\Gamma^1=\!\! \left( \begin{array}{cc}
   -\epsilon^{\alpha\beta} & 0 \\
   0 & -\epsilon_{\dot\alpha\dot\beta} \\
 \end{array}\right)\!,~{\cal C}^2=-1,~{\cal C}^T=-{\cal C},
 ~{\cal C}\Gamma^m=({\cal C}\Gamma^m)^T
\ea
where $\epsilon^{\alpha\beta}=\epsilon_{\dot\alpha\dot\beta}=
 -\epsilon_{\alpha\beta}=-\epsilon^{\dot\alpha\dot\beta}=i\sigma^2$, the
 $\sigma$'s
have indices $\sigma^m_{~\alpha\dot\alpha},~
\bar\sigma^{m~\dot\alpha\alpha}$. The minimal spinor in $d=4$
Euclidean space is Dirac (4 independent complex components) and in
the chiral representation is written in terms of two independent
Weyl bi-spinors $\psi$ and $\bar\psi$ as We also define
\bea \Lambda \; &=& \left(
\begin{array}{c}
\lambda_\alpha\\
\bar \lambda^{\dot \alpha}
\end{array}\right)
\\
\Lambda^{\cal C} &=& \Lambda^T {\cal C} = \left(\lambda^\alpha
\,\, \bar \lambda_{\dot \alpha}\right)
\\
P_\pm &=& \frac{1}{2}\, ( 1 \pm \Gamma_5 )
\eea
The conventions for contracting bi-spinors are \ba
 \psi\phi&=&\psi^\alpha\phi_\alpha=\epsilon_{\alpha\beta}\psi^{\alpha}
 \phi^{\beta}\\
 \bar\xi\bar\eta&=&\bar\xi_{\dot\alpha}\bar\eta^{\dot\alpha}=
 \epsilon^{\dot\alpha\dot\beta}\bar\xi_{\dot\alpha}
 \bar\eta_{\dot\beta}.
\ea The deformation of superspace can be rewritten as \be
 \{ \theta^\alpha,\theta^\beta\}=
 C^{\alpha\beta}=\frac12(\sigma_{\mu\nu})^{\alpha\beta} \mathbf{C}^{\mu\nu}
 \label{def}
\ee here $\sigma_{\mu\nu}=\frac 14
(\sigma_\mu\bar\sigma_\nu-\sigma_\nu\bar\sigma_\mu)$ are
anti-seldual as is $\mathbf{C}^{\mu\nu}$. Recalling that
$\theta_\alpha$ is the chiral component of the 4 component Dirac
spinor $\Theta$ we can rewrite (\ref{uno}) as \be
 \{ P_+\Theta,\Theta^\mathcal{C} P_+\}=\frac 14 P_+\Gamma_{\mu\nu}\mathbf{C}^{\mu\nu}
 \label{def4}
\ee where $\Gamma_{\mu\nu}=\frac12[\Gamma_\mu,\Gamma_\nu]$.
 The
relations (\ref{uno}) can be stated as \be
  \{ \Theta,\Theta^\mathcal{C} \}=\frac 14 P_+\Gamma_{\mu\nu}\mathbf{C}^{\mu\nu}
\ee

~

\noindent\underline{Representation for the $d=4 \to  d=3$
dimensional reduction}

\vspace{0.3 cm}

In section 4, in order to implement the dimensional reduction from
$d=4$ to $d=3$ space-time dimensions we use a Gamma matrices
representation where

\bea
 && \Gamma^i = \left(i\sigma^i\!\!
 \otimes \sigma^3 \right) = \left(
 \!\!\begin{array}{cc}
 \gamma^i & 0\\
 0 & -\gamma^i
 \end{array}\!\!\right) \, , \;\;\; {\rm for}\;\;i=1,2,3
 \\
 && \Gamma^4 =\left(i I \otimes \sigma^1 \right)
  =\left(
 \begin{array}{cc}
 0& iI\\
 iI & 0
 \end{array}\right) \, , \;\;
 \Gamma^5 =\left(I \otimes - \sigma^2 \right)
 =\left(\!\!
 \begin{array}{cc}
 0& iI\\
 -iI & 0
 \end{array}\right)
 \\
  && {\cal C} \,=\left(i\sigma^2\!\! \otimes  \sigma^3 \right) =\left(
 \!\!\begin{array}{cc}
 {\cal C}_3 & 0 \\
 0 & -{\cal C}_3 \!\!
 \end{array}\right) \, , \;\;\;
P_\pm = \frac{1}{2}\left(
\begin{array}{cc}
I & \pm iI\\
\mp i I & I
\end{array}\right)
\eea Here $\gamma^i$ can be identified with $d=3$ gamma matrices
which can be chosen as the Pauli matrices, $\gamma^i=i\sigma^i$,
this leading to \be
 {\cal C}_3 =\gamma^2
\ee

\end{document}